\documentclass[journal]{IEEEtran}
 
\usepackage[utf8]{inputenc}

\usepackage{times,enumerate} 
\usepackage[usenames,dvipsnames,svgnames,table]{xcolor}

\usepackage{graphicx}
\usepackage[noadjust]{cite}

\usepackage{rotating}
\usepackage{csquotes}
\usepackage{amsmath}
\usepackage{bm}
\usepackage{tabularx}
\usepackage{stfloats}
\usepackage{threeparttable}
\usepackage{url}
\usepackage{soul}
\usepackage{threeparttable}
\soulregister\cite7
\soulregister\ref7
\soulregister\pageref7
\usepackage{amssymb}

\usepackage{footnote}
\usepackage{colortbl}
\usepackage{soul}
\usepackage{multirow}
\usepackage{pifont}
\usepackage{algorithmic}
\usepackage{booktabs,xcolor,colortbl}
\usepackage{color}
\usepackage{alltt}
\usepackage{hyperref}
\usepackage{enumerate}
\usepackage{siunitx}
\usepackage{epstopdf}
\usepackage{bbding}
\usepackage{pbox}
 \usepackage{amsmath}  
 \usepackage{amssymb}   
 \usepackage{amsthm}
 \usepackage[dvipsnames]{xcolor}

\newcommand{\rev}[1]{\textcolor{black}{#1}}

\def\undb#1{\mbox{\bf{#1}}}

\hypersetup{
     colorlinks = true,
     linkcolor = black,
     anchorcolor = black,
     citecolor = black,
     filecolor = black,
     urlcolor = black
     }

\begin{document}

\title{Self-Healing Secure Blockchain Framework\\ in Microgrids}

\author{Suman Rath, \textit{Graduate Student Member, IEEE}, Lam Duc Nguyen, \textit{Member, IEEE}, \\ Subham Sahoo, \textit{Senior Member, IEEE} and Petar Popovski, \textit{Fellow, IEEE} 
\thanks{
This work is partly supported by the Baltic-Nordic
Energy Research programme via Next-uGrid project and partly by the Villum Investigator Grant WATER from the Velux Foundation, Denmark.

Suman Rath is with the Department of Computer Science and Engineering, University of Nevada, Reno, NV 89557, USA (e-mail: srath@nevada.unr.edu)}
\thanks{Lam Duc Nguyen is with Software Systems Research Group, CSIRO's Data61, Sydney 2015, Australia. (e-mail: lam.nguyen@data61.csiro.au)} 
\thanks{Subham Sahoo is with the Department of Energy, Aalborg University, 9220 Aalborg, Denmark (e-mail: sssa@energy.aau.dk).}
\thanks{Petar Popovski is with the Department of Electronic Systems, Aalborg University, 9220 Aalborg, Denmark. (e-mail:petarp@es.aau.dk)}}
 

\maketitle

\begin{abstract}

Blockchain has recently been depicted as a secure protocol for information exchange in cyber-physical microgrids. However, it is still found vulnerable to consensus manipulation attacks. These \textit{stealth} attacks are often difficult to detect as they use kernel-level access to mask their actions. 
In this paper, we firstly build a trusted and secured peer-to-peer network mechanism for physical DC microgrids' validation of transactions over Distributed Ledger.
Secondly, we leverage from a {physics-informed} approach for detecting malware-infected nodes and then recovering from \textit{stealth} attacks using a self-healing recovery scheme augmented into {the} microgrid Blockchain network. 
This scheme allows compromised nodes to adapt to a reconstructed trustworthy signal in a multi-hop manner using corresponding measurements from the reliable nodes in the network. \rev{Additionally, recognizing the possible threat of denial-of-service attacks and random time delays
(where information sharing via communication channels is blocked), we also integrate a model-free predictive controller with the proposed system that can locally reconstruct an expected version of the attacked/delayed signals.} This supplements the capabilities of Blockchain, enabling it to detect and mitigate consensus manipulation attempts\rev{, and network latencies}.
\end{abstract}


\IEEEpeerreviewmaketitle

\section{Introduction}

\IEEEPARstart{R}{ecent} attempts to develop more efficient DC microgrid systems can be attributed to their inherent cyber-physical capabilities that allow the smooth integration of renewable energy sources (RESs), multiple electronic devices (loads) and a variety of storage devices in both, the autonomous as well as the grid-connected mode of operation. Such systems can have three different types of control structures - centralized, distributed, and decentralized. The distributed control framework is generally preferred over the other control structures as it is resilient to single-point failures, thus enabling more reliability and better scalability \cite{sahoo2018stealth}.
Moreover, distributed control structures have a higher degree of tolerance to unavoidable cyber issues like network latency, packet losses, and communication link failures. However, these control structures can not be considered fully \textit{reliable} as the distributed framework can only enable partial information availability (limited to neighboring units), which in turn, makes the microgrid vulnerable to unauthorized manipulation of sensors and actuators  \cite{sahoo2018stealth}. Since cyber-physical DC microgrids are integral components of mission-critical systems (e.g., electric aircraft, hospitals, military bases, etc.), it is essential to protect them from adversarial cyber attacks.

Networked microgrids, which may function either in the independent mode or in collaboration with the rest of the network, can complicate the distribution system's operation.
Conventionally, through supervisory control and data acquisition (SCADA), microgrids are often controlled in a centralized manner \cite{jestpe}. In particular, the  network data is gathered, recorded, and then processed at a centralized server. Considering the rapidly growing size and speed of the measurement data, collecting and processing such a wide-range network in real-time at a single server leads to overload for the centralized node and a single point of failure problem.
\rev{Centralized networks would also be vulnerable to single-point attacks as hackers would only need to target the central host computer server to compromise the entire network, as this is the single point that issues control commands to all the other nodes \cite{deer2022centralized}.}
In contrast, attacking a distributed network is more challenging for attackers, especially for the power grid that is distributed over a geographically wide-range area. Hence, the computation and the trust are distributed over the network, which can be useful for achieving a more secure power network.

In this regard, as  the energy sector needs significant time reduction for the management of economic transactions and the possibility of getting rid of third-party authorities \cite{mollah2020Blockchain}, the concept of \textit{transactive energy} has appeared as one of the most interesting technologies since the 2017 Gartner Hype cycle \cite{TopTrend97:online}.
There is a need to switch from a centralized energy system with intermediary components to a decentralized system that can detach the related financial transactions from the centralized energy control unit. For example, the authors in \cite{di2018technical} introduced an approach using Blockchain technology for handling loss allocation and a timing mechanism for transacting intended energy exchanges and losses.

A type of Distributed Ledger Technology (DLT) known as Blockchain consists of blocks, which are lists of data items that are constantly growing. Transactions of all kinds, including those involving money, energy, transportation, data, logic, and even programs, can be recorded in blocks. Blocks are tied together in chronological order, timestamped, unchangeable, and verifiable. Every DLT node keeps a copy of the ledger, and any changes to the blocks' content or order are quickly detected by a check of the authenticity of the blocks. Among the concerned untrusted parties, Blockchain encourages immutable and transparent information sharing \cite{9324804}. DLTs are viewed as a crucial enabler for trustworthy and dependable distributed observation systems in addition to their function in financial transactions. DLTs' authentication procedure depends on network consensus across numerous nodes. Blockchain technology is a noteworthy invention that can be applied in various microgrid domains.

\rev{Recently, Blockchain has become more popular in the integration with microgrids. There are several prominent applications of Blockchain technologies in microgrids, e.g., peer-to-peer energy trading \cite{wu2022p2p}, energy exchange \cite{laszka2018transax}, electric vehicle charging \cite{umoren2020blockchain}. Besides, several start-ups and companies have applied Blockchain technologies in microgrids system for managing and sharing energy as well as building innovative products, e.g., the energy market of PowerLedger \cite{Powerled18:online}, a decentralized data exchange platform for energy sector GridSingularity \cite{GridSing14:online}. More applications of Blockchain in microgrids can be found in \cite{goranovic2017blockchain}.}
The authors in \cite{gaybullaev2021efficient} have discussed a Blockchain-based scheme for secure energy transactions. In conventional Blockchain-based schemes, the trustworthiness of each block is decided through a consensus-based mechanism where the majority of nodes in the system must agree on its validity. Such schemes are often vulnerable to 51$\%$ attacks \cite{aggarwal2021review}. In general, 51$\%$ attacks refer to an attack strategy where the hacker can access the system's internal mining capabilities and use this access to mine compromised blocks. Further, this access is also used to manipulate the consensus-based validation strategy by forcing individual nodes to stop the confirmation of authentic blocks and make them validate fabricated blocks generated by the attacker itself. 51$\%$ attacks can be executed using stealthy, kernel-level malware like rootkits \cite{rath2022behind}.

Although the Blockchain\footnote{Throughout this article, the words DLT and \emph{Blockchain} will be used interchangeably. Blockchains are a particular sort of DLT where each node keeps a copy of the ledger, and chains of blocks are made up of digital pieces of information called transactions.} is being used in many power electronic applications, it is also essential to evaluate whether Blockchain could be a feasible tool for exchanging voltage/frequency control signals in the microgrid test system. This evaluation is essential because, even though the Blockchain is an essential tool for cyber-secure control, the time delay created as a consequence of mining new blocks may have a detrimental effect on microgrid stability \cite{yao2021distributed}. Mahmud \textit{et al.} in \cite{mahmud2021Blockchain} have proposed the use of Blockchain for control in microgrids with distributed energy resources and presented results showing that the proposed framework can perform well even under time delays. However, the current studies have not addressed how to prevent stealth \& hijacking attacks in their proposed control mechanism.

\rev{The conventional control framework in cyber-physical microgrids critically relies on a centralized database and communication channels, both of which are vulnerable to adversaries to manipulate the system operation. An effective solution for countermeasures is the detection and mitigation of such threats. However, mitigation techniques may be vulnerable to adversarial attacks designed with knowledge of the system dynamics (e.g., adversarial attacks on learning-based anomaly detectors, reward poisoning attacks against reinforcement learning, etc.). Hence, a significantly better approach is the prevention of these threats in the first stage itself. Since the DLT technology offers great potential due to its advanced data protection and attack prevention capabilities based on the natural security strength of the cryptographic system and consensus mechanism \cite{soret2021learning}, we propose to augment these salient features of the DLT into the dynamic physical properties of microgrids to enhance its resiliency against cyber-attacks {(along with physics-informed attack detection metrics to identify {any type of consensus-manipulating attack vector that tries to inject false data into the network)}}. To recover from such attacks upon detection, we introduce a self-healing recovery mechanism for Blockchain-based microgrids that: i) provides a transparency and immutability microgrids system; ii) addresses the \textit{garbage-in garbage-out} problem of the current Blockchain-based systems.}

To summarize the key features, the contributions of this paper are as follows:

\begin{itemize}
    \item  Firstly, we formulate a general model of a DLT-based transactive DC microgrid system, including infrastructure and security mechanisms.
   
    \item {Secondly, we evaluate the response of the general Blockchain model to stealth attacks and propose the addition of new physics-informed detection metrics based on the secondary control dynamics of each node to identify such attacks and alleviate the detection accuracy of DLT.}
    
    \item {Finally, we also present a self-healing recovery mechanism in the form of a reconstructed signal to compensate for any compromised data/signal at any node. This recovery mechanism is independent of the physical topology of multiple converters, and will always provide system resiliency, provided at least one node in the network is trustworthy. \rev{The presented strategy is resilient against several attack variants and potential network issues including FDIA, denial-of-service (DoS), and random communication delays.}}
    
\end{itemize}

{The remainder of this paper is organized as follows. In Section II, we introduce the background knowledge of DLTs/Blockchain and the integration of Blockchain into microgrids. Section III highlights the operating principles and security issues in cyber-physical microgrids. Section IV presents the Blockchain-based strategy adopted for cyber-secure control in the DC microgrid. The effectiveness and robustness of the proposed strategy to illustrate its resilient behavior under different attack scenarios and physical topologies of microgrids are presented in Section V. \rev{Experimental results demonstrating the robustness of the proposed self-healing strategy are presented in Section VI.} Finally, Section VII presents the conclusion of the paper.}

\section{Background Knowledge}

\subsection{DLTs/Blockchain Concept}
\begin{figure}
    \centering
    \includegraphics[width=0.8\linewidth]{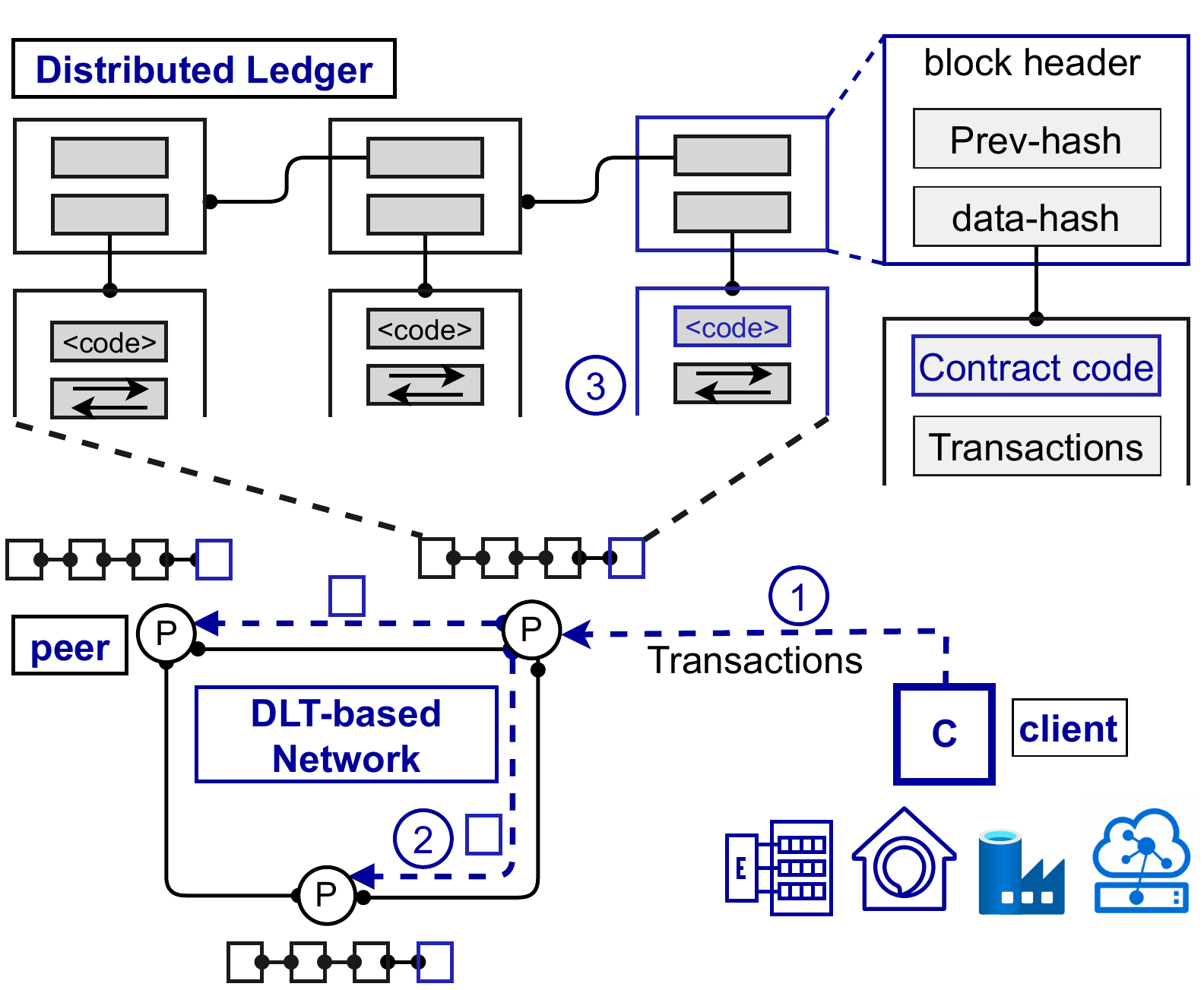}
    \caption{Schematic of a DLT-based system. }
    \label{fig:dlt-background}
\end{figure}

A Distributed Ledger Technology (DLT) system provides a distributed, tamper-proof ledger that is spread throughout a network of interacting nodes, which share a common initial block of data known as the genesis block \cite{nguyen2022blockchain}. When publishing data to the ledger, each node adds data formatted in the form of transactions in a block that also contains a pointer to its preceding block. This produces a chain of blocks, or "Blockchain," referred to as the Blockchain. To receive a reward, a block that is generated by a particular node often needs to solve a mathematical crypto-puzzle \cite{courtois2014optimizing} and provide the result as evidence of its effort. Mining is the name of this process. The network's overall computing or mining power changes the crypto-level puzzles of difficulty.
Since every DLT node in the network maintains a copy of all committed transactions in the ledger, every transaction recorded in the distributed ledger is practically unchangeable \cite{nguyen2020witness}.
Additionally, the integrity of the data blocks in the DLTs is ensured by cryptographic methods, including hash functions, asymmetric encryption algorithms, and digital signatures. As a result, the DLTs can guarantee transaction non-repudiation. Each transaction is also timestamped historically and given a unique ID, allowing each user to be assigned to it. The life cycle of a transaction in DLT-based networks is explained in detail in Fig. \ref{fig:dlt-background}. Blockchain clients, e.g., physical devices and home alliances, generate or collect data, and transform it into Blockchain transaction format.
The transactions are subsequently put into blocks, which are validated by peers via the mining process. In the distributed ledger, the transactions are finally immutably recorded and accessible to clients. After the transactions are recorded in the distributed ledger, they can be queried by clients.

\subsection{Cyber Attacks on Microgrids}

Blockchain is regarded as a typical distributed data storage system and encompasses a number of other technologies as well, including decentralization, distributed consensus procedures, and cryptography. Blockchain has been studied in both research and applications due to the benefit of creating safe, dependable, and decentralized autonomous ecosystems for a variety of scenarios \cite{lin2017survey}. Blockchain is suited for applications involving the security protection of cyber-physical systems (CPSs) because it is a novel and fundamental technical feature.

Recent cyber attacks against critical physical systems, for example, the attacks on the Ukrainian power grid in 2015-2016, have motivated multiple studies on security and privacy aspects for cyber-physical systems, particularly the power grid \cite{deng2017ccpa}. 
The work in
\cite{xie2011integrity} demonstrated that\textit{ false data injection} (FDI) could inject data measurements to induce error in the operation of the power system. A malicious attacker can also affect forecast systems, which are used to plan the operation and activities of power systems by exploiting vulnerabilities of artificial intelligence models \cite{chen2019exploiting}. 
In general, FDI attacks need to gather information about the current state of the system or the models used for making decisions, for example, the state of the system, topology structure, or machine learning prediction models \cite{barreto2018impact}. 

Besides, the attacker also exploits the lack of security mechanisms of embedded IoT sensors \cite{rath2022behind} to affect the operation of power grid systems.
The target of adversaries could be embedded sensors such as environmental sensors, smart meters, appliances, or end-user systems to affect the power systems. The authors in \cite{huang2019not} introduced a novel form of attack named manipulation of demand via IoT and documented that if an attacker compromised thousands of nodes, they could cause various problems to the power grid, including line failures, frequency instabilities, increased operating costs. In another aspect, the adversaries can compromise IoT devices responsible for communicating or exchanging data and payments with other power systems by changing their bids \cite{mengis2017data}. 
In the scope of this work, we propose to augment the physical properties, dynamics, and principles of microgrid operation into Blockchain-based attack detection and mitigation.

\section{Cyber-Physical Microgrids}
\subsection{Operating Principle}

\begin{figure}
    \centering
    \includegraphics[width=0.9\linewidth]{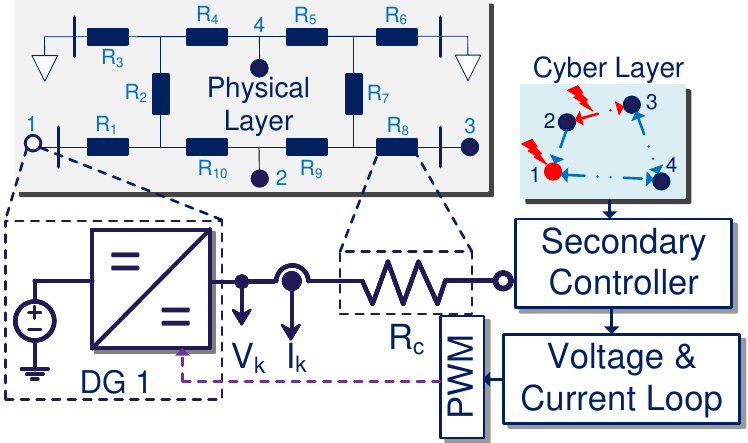}
    \caption{Networked DC microgrid with $N$ = 4 distributed generations (DGs) operating with a cooperative ring-based cyber graph.}
    \label{fig:sys_model}
\end{figure}

A single-line diagram depicting a cyber-physical DC microgrid network with $N=4$ sources and corresponding DC/DC buck converters is shown in Fig. \ref{fig:sys_model}. Each source with the corresponding converter is called an \textit{agent}. Further, the agents are connected to each other via transmission lines. Besides the physical interconnections, the agents are also connected to one another through a communication network that aids in the exchange of information between themselves. The information received from the communication network is fed as input to the local controller associated with each agent. As shown in Fig. \ref{fig:sys_model}, the local control framework consists of voltage and current controllers for the management of the DC/DC converter. Apart from that, the secondary control framework consists of an average voltage regulator for facilitating the regulation of global voltage, and a current regulator to achieve proportionate load sharing through the imposition of voltage offsets from each of the layers.


An undirected, interconnected graph of cyber elements is shown in Fig. \ref{fig:sys_model}. In this graph, the nodes represent the agents, and are denoted by $\textbf{x} = \{{x}_1, {x}_2, \ldots, {x}_N\}$). The nodes are connected to each other using edges through an associated adjacency matrix, ${\mathbf A_\text{G}} = [{a}_{kj}]\in{R^{N\times{N}}}$, where the communication weight (represented by $a_{kj}$, i.e., from node $j$ to node $k$) is formulated as: $a_{kj} >$ 0, if ($\psi_k$, $\psi_j$) $\in$ $\mathbf{E}$, where $\mathbf{E}$ represents an edge connecting two different nodes, with $\psi_k$ and $\psi_j$ representing a local node and its neighboring node, respectively. If the cyber link connecting $\psi_k$ and $\psi_j$ is absent, $a_{kj}$ = 0. In this framework, an agent (say at $\psi_k$ node) shares local voltage and current measurements with its neighbors $N_{k} = \{ {j} \ | \ ({\psi}_j, {\psi}_k) \in \mathbf{E} \} $. The matrix showing input and output information can be represented as $\mathbf{D}_\text{in} = \texttt{diag}\{{d}_k^{in}\}$ and ${\mathbf{D}_\text{out}} = \texttt{diag}\{{d}_k^{out}\}$ respectively, where, ${d}_k^{in}=\sum_{j\in N_{k}}a_{kj}$ and ${d}_k^{out}=\sum_{i\in N_{j}}a_{jk}$. Combining the transmitted and received information, a single Laplacian matrix can be obtained, which is denoted by $\mathbf{L}$ = [${l}_{kj}$]. The elements of the matrix are represented by $l_{kj}$ and obtained using $\mathbf{L}$ = ${\mathbf{D}_\text{in}} - {\mathbf{A}_\text{G}}$.

\begin{figure*}
    \centering
    \includegraphics[width=.75\linewidth]{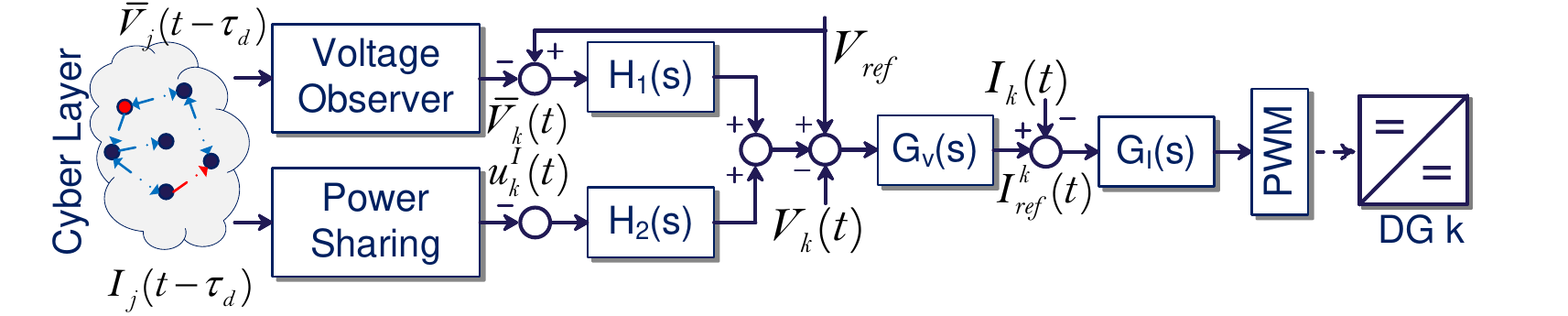}
    \caption{Distributed controller for microgrids under consideration in $k^{th}$ DG.}
    \label{fig:control}
\end{figure*}

As previously mentioned, the role of the cooperative control framework is to achieve average global regulation and proportional sharing of load current. To fulfill these objectives, a reference value of the voltage setpoint is determined through the use of two voltage correction terms, as defined below:
\begin{align}
 \Delta V_{1k}(t) &  =K_{P}^{H_1}(V_ {\rm dcref} - \bar{V}_i (t))
+K_{I}^{H_1}\int(V_ { \rm ref}- \bar{V}_k (t))dt
\end{align}
\begin{equation}
\Delta V_{2i}(t)=K_{P}^{H_2}\delta _{k}{ (t)}+K_{I}^{H_2}\int\delta _{k}{ (t)}dt
\end{equation}
where $\bar{V}_{i}$ represents the estimated average value of the voltage at the $i^\text{th}$ agent; ${V}_{\rm ref}$ represents the nominal value of voltage; $\delta _{k}$ represents the current mismatch error (in (4)) for the $k^\text{th}$ agent between the local per-unit output current and the neighbors’ per-unit output current. The output from the voltage observer and power-sharing controller in Fig. \ref{fig:control} can be mathematically depicted as:
\begin{equation}\label{enq:e1}
\bar{V}_{k}(t)=V_{{k}}(t)+\int \sum _{j\in N_{k}}a_{kj}(\bar{V}_{j}(t-\tau )-\bar{V}_{k}(t-\tau ))dt
\end{equation}
\begin{equation}\label{enq:e2}
\delta _{k}{ (t)}=\sum _{j\in N_{k}}ca_{kj}\left (  \frac{I_{{j}}(t-\tau)}{I_{{j}}^{max}}-\frac{I_{{k}}(t-\tau )}{I_{{k}}^{max}}\right )
\end{equation}
where, $\tau$ is the communication delay between the $k^\text{th}$ \& the $j^\text{th}$ agent and $c$ represents the coupling gain. Further, $I_{{k}}$ and $I_{{j}}$, $I_{{k}}^{max}$ and $I_{{j}}^{max}$ represent the measured and the maximum output currents for the $k^\text{th}$ 
 and the $j^\text{th}$ agent, respectively.

The local reference value of voltage $V_{k}^{*}$ for the $k^\text{th}$ agent, determined by using the two voltage correction terms as depicted in (1)-(2) can be given by:
\begin{equation}
V_{k}^{*}(t)=V_ { \rm ref}+\Delta V_{1k}(t)+\Delta V_{2k}(t).
\end{equation}

For a well-connected cyber graph in a networked DC microgrid, based on the cooperative consensus algorithm, the global control objectives can be given by:
\begin{equation}\label{enq:e6}
\lim_{k\rightarrow \infty }\bar{V}_{k}(t)=V_{\rm ref},\quad \lim_{k\rightarrow \infty }\delta _{k}(t)=0.\quad\forall k\in N
\end{equation}
\begin{table}
	\renewcommand{\arraystretch}{1.3}
	\caption{Stealth Attacks in DC Microgrids in \cite{sahoo2018stealth} and \cite{sahoo2019detection}}
	\label{tab:1}
	\centering
	\begin{tabular}{c|c}
		\hline
		Affected Counterparts & Modeling \\
		\hline
		\hline
		
		Voltage \cite{sahoo2018stealth} & $\undb{W} {x}^V_{attack}$ = 0 \\
		\hline
		
		Current \cite{sahoo2019detection} & $\undb{W} {x}^I_{attack}$ = 0\\
		\hline
	\end{tabular}
\end{table}

\subsection{Cybersecurity Issues}
Fig. \ref{fig:sys_model} shows how malicious attackers may inject false data into the sensors, actuators, controllers and other cyber devices in the DC microgrid system to disrupt the objectives depicted in (\ref{enq:e6}). These attacks may also be performed in a coordinated fashion in order to deceive microgrid system operators and hide the actions of the attack vector by adding additional inputs in (\ref{enq:e1})-(\ref{enq:e2}), as given by:
    \begin{eqnarray}
\undb{u}^{a}(t) = \undb{L} \undb{x}(t) +  \undb{W} \undb{x}_{attack} \label{12}    \end{eqnarray} 
where, $\undb{u}^a$ is the vector representation of the manipulated control input $u^a_k$ = \{$u^{Va}_k, u^{Ia}_k$\}. Further, $\undb{x}_{attack}$ and $\undb{x} = \{ \bar{\undb{V}}, \undb{I}\}$ represent the attack elements $x_{attack_{i}}$ = [$x^{V}_{attack_{k}}, x^{I}_{attack_{k}}]^T$ and the non-compromised measurements, respectively. Moreover, { the attack distribution matrix} $\undb{W}$ = [$w_{kj}$] represents row-stochasticity, where its elements can be given by:
\begin{eqnarray}
w_{kj} = {\begin{cases}
	\frac{1}{N_k + 1}, \ j \in N_k\\
	1 - \sum_{j \epsilon N_k} w_{kj}, \ j = k \\
	0, j \not\in N_k, j \neq k
	\end{cases}} \label{14}
\end{eqnarray}
The detection strategies for both the stealth attacks (in Table \ref{tab:1}) are provided in Table \ref{tab:2}. More details on its formulation can be referred from \cite{sahoo2018stealth} and \cite{sahoo2019detection}. As it can be seen in Table \ref{tab:2} that since the detection strategies primarily depend on local as well as neighboring measurements, this leaves a tactical opportunity for an attacker to infiltrate the communicated measurements used for the design of the detection strategies. As a result, such consecutive intrusion may affect the detectability of these attacks when the criteria in Table \ref{tab:2} are not met. 
\begin{figure}
    \centering
    \includegraphics[width=0.75\linewidth]{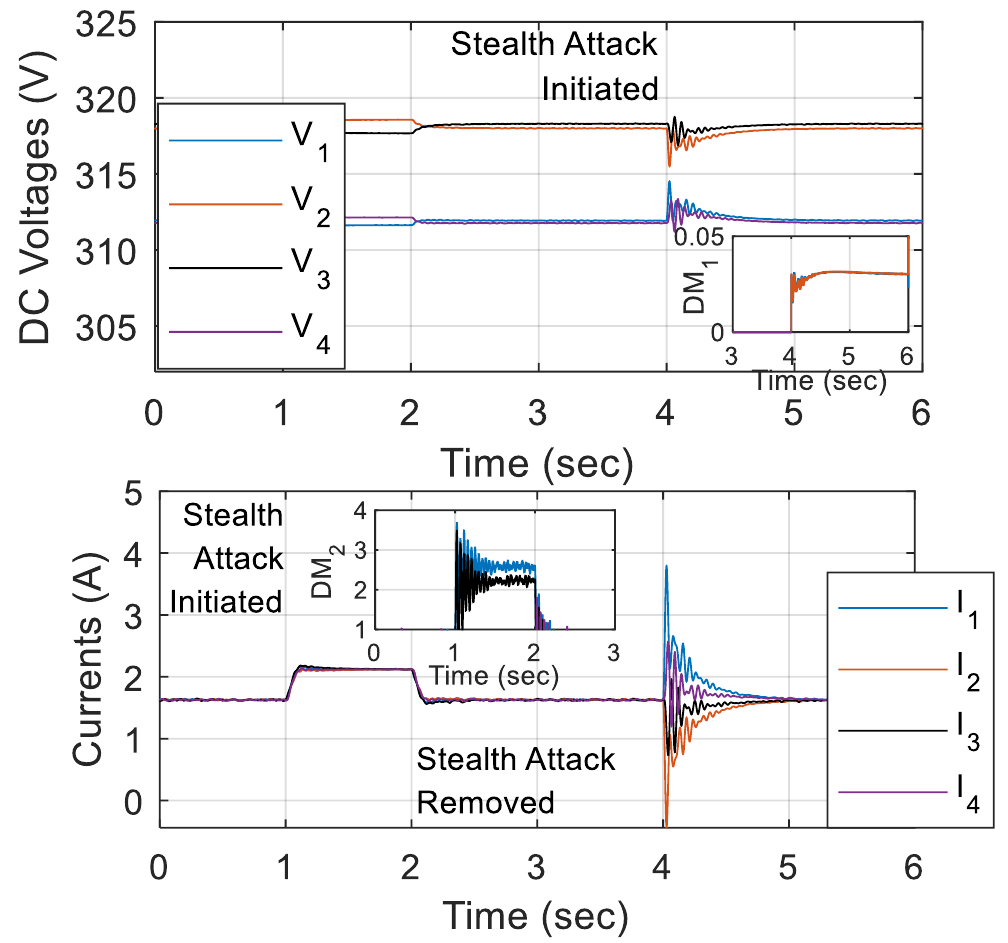}
    \caption{Performance of detection strategies in response to stealth attack on current and voltages at t = 1-2 sec and 4 sec, respectively -- the detection strategies can further be manipulated.}
    \label{fig:12}
\end{figure}

For more clarity, a case study is performed in a DC microgrid system with $N$ = 4 agents, as shown in Fig. \ref{fig:12} to demonstrate the performance of detection criteria for a stealth attack on voltages and currents, respectively. In Fig. \ref{fig:12}, after the stealth attack is initiated on currents, it can be seen that the output currents of each agent are anyway shared equally. However, the detection mechanism for current measurements suggests that $DM^1_2$ and $DM^3_2$ are positive, which indicate compromised current measurements for agent I and III. Further at t = 4 sec, a balanced set of attack elements \{-15, 0, 15, 0\} V are introduced in the voltage control input values (as shown in Fig. \ref{fig:12}) on the basis of the attack strategy depicted in \cite{sahoo2018stealth}. Despite the presence of attack elements, the voltages return back to pre-attack set points. However, the corresponding detection metric $DM_1$ for agents I and II immediately goes positive. Under these circumstances, if the communicated measurements $x_j (t)$ in (1) are manipulated by the cyber-attacker with considerable knowledge of the detection strategy, there is a possibility that the cyber attack alarm might be disinformed. This mandates a secure mechanism of information exchange between each agent so that the communicated variables used in the proposed detection approach remain uncompromised. Following an accurate scanning of the compromised measurement, the corresponding countermeasures can be applied without any false alarms. 
\begin{table}
	\renewcommand{\arraystretch}{1.3}
	\caption{Detection Criteria for Stealth Attacks in Table \ref{tab:1}}
	\label{tab:2}
	\centering
\begin{threeparttable}
	\begin{tabular}{c|c|c}
		\hline
		Stealth Attack & Detection Criteria for $k^{th}$ Agent & Terminology \\
		\hline
		\hline
		
		Voltage \cite{sahoo2018stealth} & \begin{tabular}{@{}c@{}} $h_k$\tnote{1} $[\sum_{j \epsilon N_k} a_{kj} (\Delta V_{1_{j}} - \Delta V_{1_{k}})]$\\ $[\sum_{j \epsilon N_i} a_{kj} (\Delta V_{1_{j}} + \Delta V_{1_{k}})] >$  $\Upsilon_1$ \end{tabular}& $DM^k_1$ \\
		\hline
		
		Current \cite{sahoo2019detection} & \begin{tabular}{@{}c@{}} $c_k [\sum_{j \epsilon N_k} a_{kj} (I^{j}_{in_{ref}} - I^{k}_{in_{ref}})]$\\ $[\sum_{j \epsilon N_k} a_{kj} (I^{j}_{in_{ref}} + I^{k}_{in_{ref}})] >$  $\Upsilon_2$ \end{tabular}\tnote{2} & $DM^k_2$\\
		\hline
	\end{tabular}
  \begin{tablenotes}
	\item[1] $h_k$ is a positive quantity used for $i^{th}$ agent.
	\item[2] $c_k$ is another positive quantity. $I^{k}_{in_{ref}}$ represents the value of input current reference for the $k^{th}$ agent.
\end{tablenotes}
\end{threeparttable}
\end{table}

\section{Blockchain for Security of Microgrids}

In this section, we present our self-healing Blockchain-based communication model for the cyber-physical microgrid shown in Fig. \ref{fig:sys_model}. We propose a distributed DLT, where the nodes can communicate and collaborate to identify attacked counterpart(s) in a reliable and trustworthy manner. 
The DLT-based network consists of nodes connected via the P2P model. Basically, nodes connect using a unique address and use the gossip protocol to exchange network information such as blocks, transactions, and addresses. In DLT-based networks, there are specific nodes called miners responsible for extending the Blockchain by creating new blocks \cite{andoni2019Blockchain}. In the scope of this research, each node aggregates sensor measurement values (in its area) and analyzes them. Further, each of them estimates the local update and exchanges signal values with the other nodes, engages in distributed consensus procedures, and publishes to the distributed ledger in its local memory.  

\subsection{System Model}

We consider the DLT-based microgrid system with $M$ communication channels and $N$ physical agents. The system states have already been defined as $x_i(t)$ = $\{V_k(t), I_k(t), Det V_k(t-1), Det I_k(t-1)\} $. The measurements $y_i(t)$ = $\{V_k(t), I_k(t), Det V_k(t-1), Det I_k(t-1)\} $ can be measured from every agent. The microgrid is modeled as a discrete-time linear dynamic system, as given below: 

\begin{equation}
    x_t = \undb{A}x_{t_1} + v_t
\end{equation}
\begin{equation}
    y_t = \undb{H}x_t + w_t
\end{equation}
where $A\in R^{(2N-1) \times (2N-1)}$ and $H\in R^{K \times 2N-1}$ represent the state transition and the measurement matrices respectively. Further, $v_t = [v_{1,t}, ..., v_{2N-1,t}]^T$ and $w_t = [w_{1,t}, ..., w_{K,t}]^T$ represent the process noise and the measurement noise vectors respectively. It is assumed that $w_t$ and $v_t$ are independent AWGN processes, where $w_t \sim \mathcal{N}(0, \sigma^2_wI_K)$ and $v_t \sim  (0, \sigma^2, I_{2N-1})$. 

For a secure and reliable state update among DLT-based nodes, it is essential to guarantee that the previous values of the state estimate are not modified, the physical agent measurement is non-anomalous, and the operation of the DLT nodes adheres to the predefined agreements. In case the network is attacked, the state estimate values can be recovered through the use of previous reliable state estimations that also require the protection of preceding state estimation values against tampering. 
On the one hand, in a DLT-based network, malicious adversaries can obtain illegitimate access to the system, for example, via trojan propagation, stealing the identity of involved nodes, leading the nodes to get faulty updates during the operation of the system. On the other hand, as the characteristic of DLT is completely distributed, there is no single trustworthy central entity to take control and verify whether the other nodes are reliable and safe, e.g., if the involved nodes are operating as per pre-defined roles. Hence, a distributed authentication mechanism is needed to detect misbehaving nodes.

{\subsection{Detection Model}
In a DLT-based microgrid network, even though the whole system is fully distributed, attackers can still manipulate the system dynamics if they can achieve control of the majority of the individual nodes, {which gives them access to more than half of the hashing/computational power}. This scenario {can be used to manipulate consensual query decisions to introduce false data into sensors/communication links in the system \cite{aggarwal2021review, xie2011integrity}. In certain cases, the attackers may also create a \textit{fork}, extracting specific transactions, tampering with them, and injecting them back as \texttt{True} packets.}
Since these attack {vectors} were designed to prey on the vulnerability of the conventional DLT framework, we {integrate system-specific knowledge into} the DLT setup {enabling it to recognize violations in microgrid} dynamics {to identify false data} disrupting the secondary controller synchronization among nodes. The generated data from individual nodes are formatted in Blockchain-based transactions and synchronized among them. The detection criteria as mentioned in Table \ref{tab:2} are in the form of mismatch error margins {(represented by the metric $DM^k$)} between parameter data (measurements) from the distributed nodes. When {an attack vector stealthily manipulates transactions, the detection metric $DM^k$ increases beyond the threshold value $\Upsilon$ predefined in the smart contract, enabling autonomous attack detection using the physics-informed attack detection metrics in Table \ref{tab:2}.}
%
A more elaborate explanation of the integrated setup including the self-healing process is depicted in the next subsection.}
\begin{figure}
    \centering
    \includegraphics[width=0.9\linewidth]{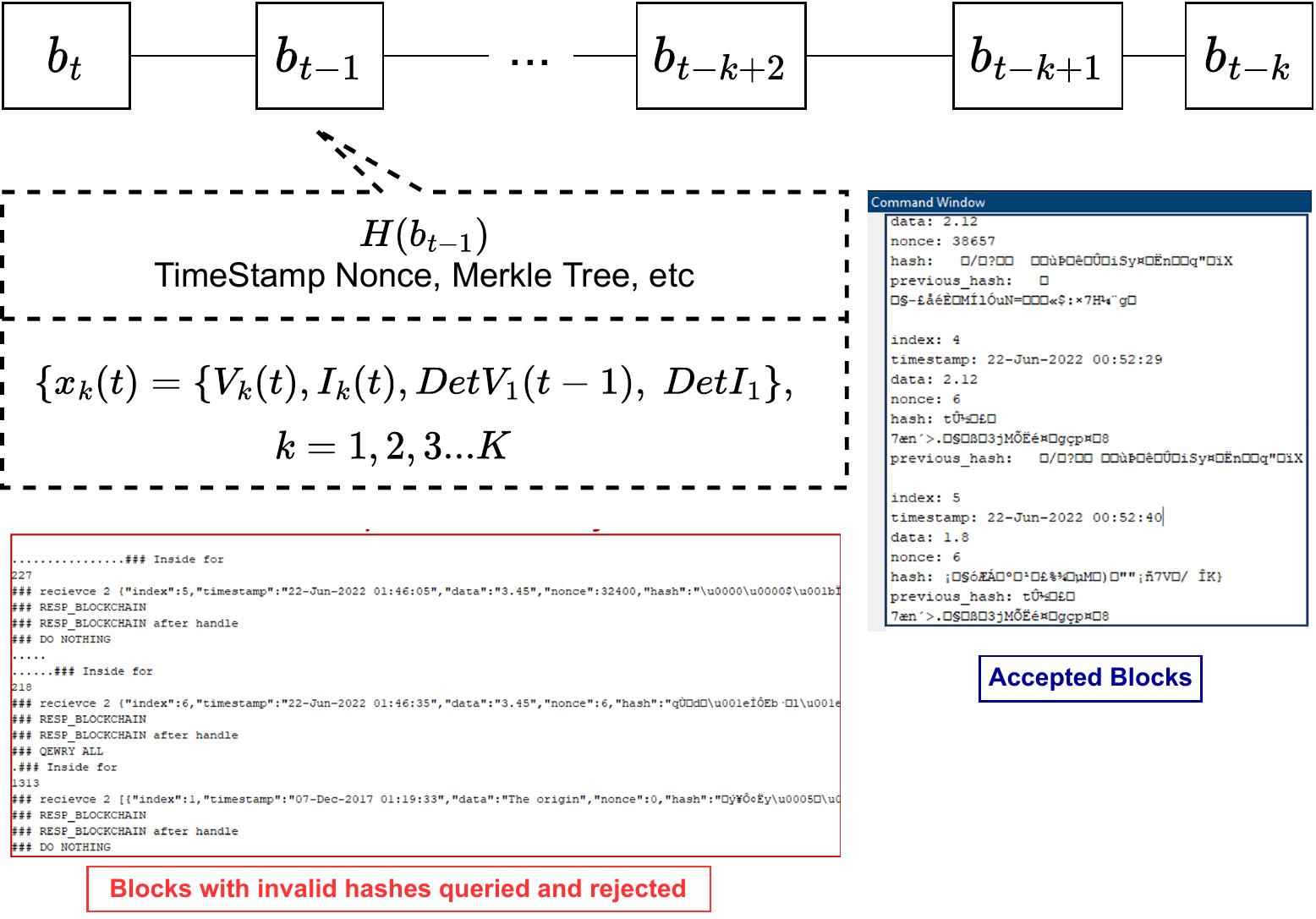}
    \caption{Distributed Ledger and structure of $b$ block at time $t$. {N}ote that $b_t$ is the block generated at time $t$ and hash of the SHA H(.) {algorithm}. The block includes a block header and body of the local state vectors in corresponding time.}
    \label{fig:Blockchain_blocks}
\end{figure}

\subsection{Self-healing recovery}

{The self-healing process designed to enable recovery of compromised nodes in the DLT-microgrid framework is part of an integrated setup including both attack detection and mitigation.} {{Any transaction performed by a node, leading to the generation} of a new block of data is broadcast to the rest of the network through a secure sharing scheme. The network uses the recipient node's ability to validate transactions using digital signatures in order to detect any malicious data in the system. This process is carried out through the determination of \textbf{\textit{events}}, when an external agent attempts {data alteration}, leading to a {modification of} the hash index associated with the detection metrics
in the corresponding block. All subsequent changes lead to further variation in the computed hash index, which is detected by the recipient node through comparison with the last trustworthy block received. In the event of a mismatch, the recipient queries the rest of the nodes for verification. As shown in Fig. \ref{fig:Blockchain_blocks}, any hash which is declared \texttt{False} by the majority of the nodes is considered to be associated with a manipulated block of data and hence, rejected by the receiver. Thus, the general anomaly detection framework is based upon a mutual trust-based structure where a block is acceptable, if and only if it satisfies the following criteria:}
\begin{equation}
    {N_a > N_r}
\end{equation}
{where, $N_a$ represents the number of approving nodes, and $N_r$ is the number of rejecting nodes. {Since} this framework is vulnerable to consensus-deceiving hijacking attacks, an {additional} set of detection criteria as depicted in Table \ref{tab:2} is {defined using smart contracts and} utilized to detect such deceptive attacks. Any \textit{event} resulting in the violation of the detection criterion in Table \ref{tab:2} leads to the labeling of the agent as a misbehaving node.}
\begin{figure}
    \centering
    \includegraphics[width=\linewidth]{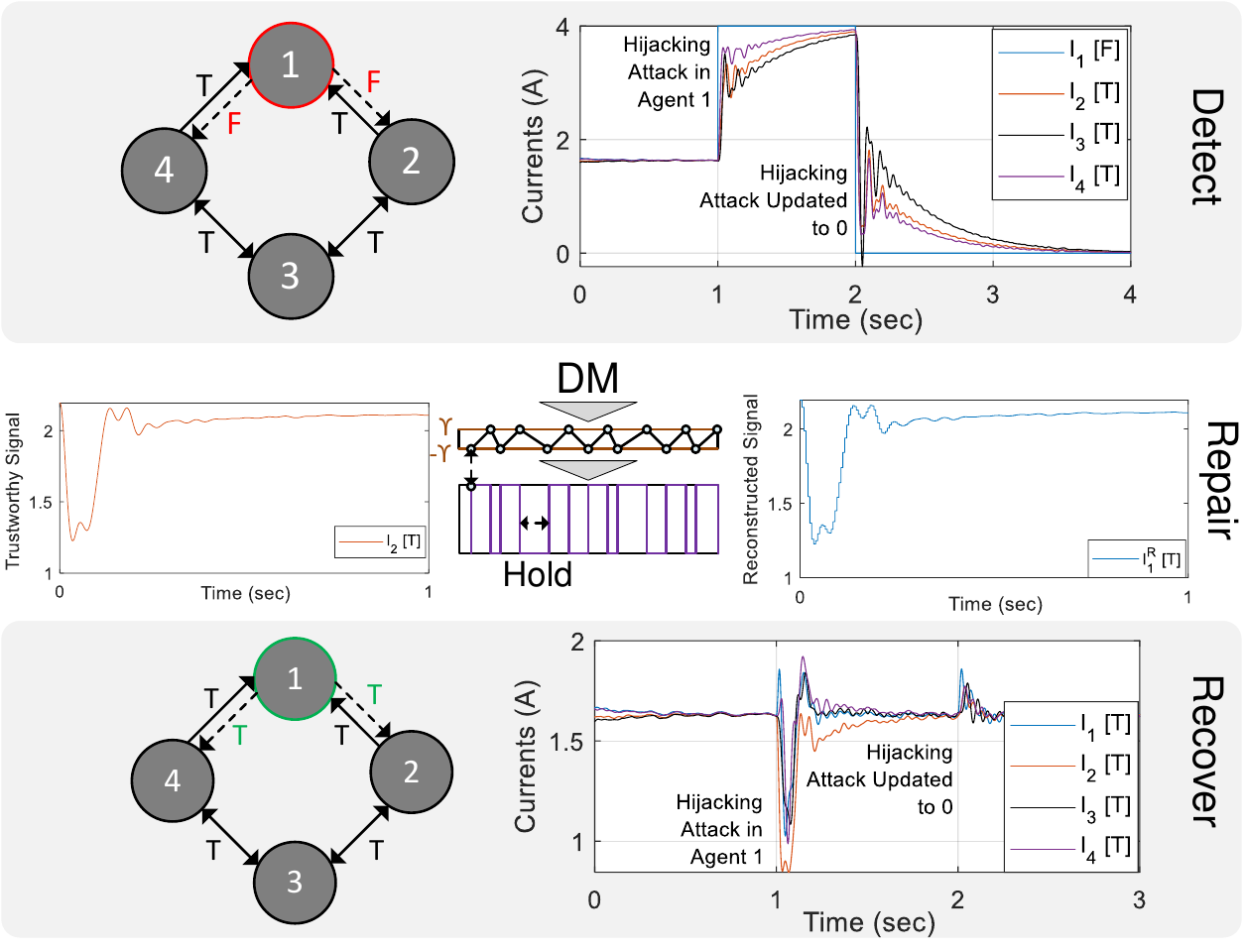}
    \caption{Self-healing recovery scheme augmented into DLTs for microgrids -- the detection metrics in Table \ref{tab:2} initiate the reconstruction process using trustworthy measurements from neighbors and then self-heal the system by transmitting back the reconstructed signal $I_1^R (t_i)$.}
    \label{fig:Self-healing_1}
\end{figure}

{Detection of a misbehaving node triggers the activation of a self-healing strategy where:
\begin{enumerate}
    \item firstly, the compromised {transaction is canceled, stopping further communication of any manipulated}
    signal to neighboring nodes
    to limit the attack propagation.
    \item secondly, the malicious {transaction is}
    not recorded in the ledger and a trustworthy version of the {manipulated} signal
    is reconstructed for each compromised node (using pseudo-anonymous values of trustworthy sensor {data}
    stored in the irreversible ledger), to preserve system stability. This has been shown in Fig. \ref{fig:Self-healing_1}.
    \item finally, this trustworthy event-driven signal is sampled from measurements of the neighboring nodes (detailed philosophy can be found in \cite{event1}-\cite{event2}) and previous measurements from the same node obtained from the ledger as per the following equations:
\end{enumerate}  }

\begin{equation}
    {\Delta V_{1_j}(t_i) = f_1 (\Delta V_{1_k}(t), \Delta V_{1_j}(t_{i-1}))}
    \label{r1}
\end{equation}
\begin{equation}
    {\Delta I_{dc_j}(t_i) = f_2 (\Delta I_{dc_k}(t), \Delta I_{dc_j}(t_{i-1}))}
    \label{r2}
\end{equation}
{where, $i$ represents the instant at which the event occurs (also called the triggering instant), and $f(\circ)$ represents the triggering function as per which the reconstruction is performed. A major role of $f(\circ)$ is to hold the input value of the signal until the next triggering event. Hence, the recipient node is localized with no communication inputs from the malicious node. Ultimately, all the manipulated blocks are substituted with event-driven, secure reconstructed blocks.} {This strategy is followed to reconstruct the manipulated signal(s) at each compromised node.}
{The presence of the terms, $\Delta V_{1_k}(t)$ and $\Delta I_{dc_k}(t)$ in (\ref{r1}) and (\ref{r2}) imply that the trustworthy reconstruction of a manipulated signal will always be possible if at least one of the nodes (here, represented as the $k^{th}$ node) is reliable (i.e., not affected by the attack). In an event where even the $k^{th}$ node is infected (i.e., 100$\%$ infection), the mitigation strategy will fail to reconstruct a trustworthy version of the signal. Hence, the resiliency scale of the proposed self-healing recovery scheme can be given by $N - 1$, with $N$ given by the number of converters in the system. In simple terms, the above sentence implies that in a system comprising of $N$ nodes, the proposed scheme will always provide secure and resilient behavior, even if there is only one trustworthy node to broadcast its information to self-heal the system.}

\begin{figure}
    \centering
    \includegraphics[width=0.65\linewidth]{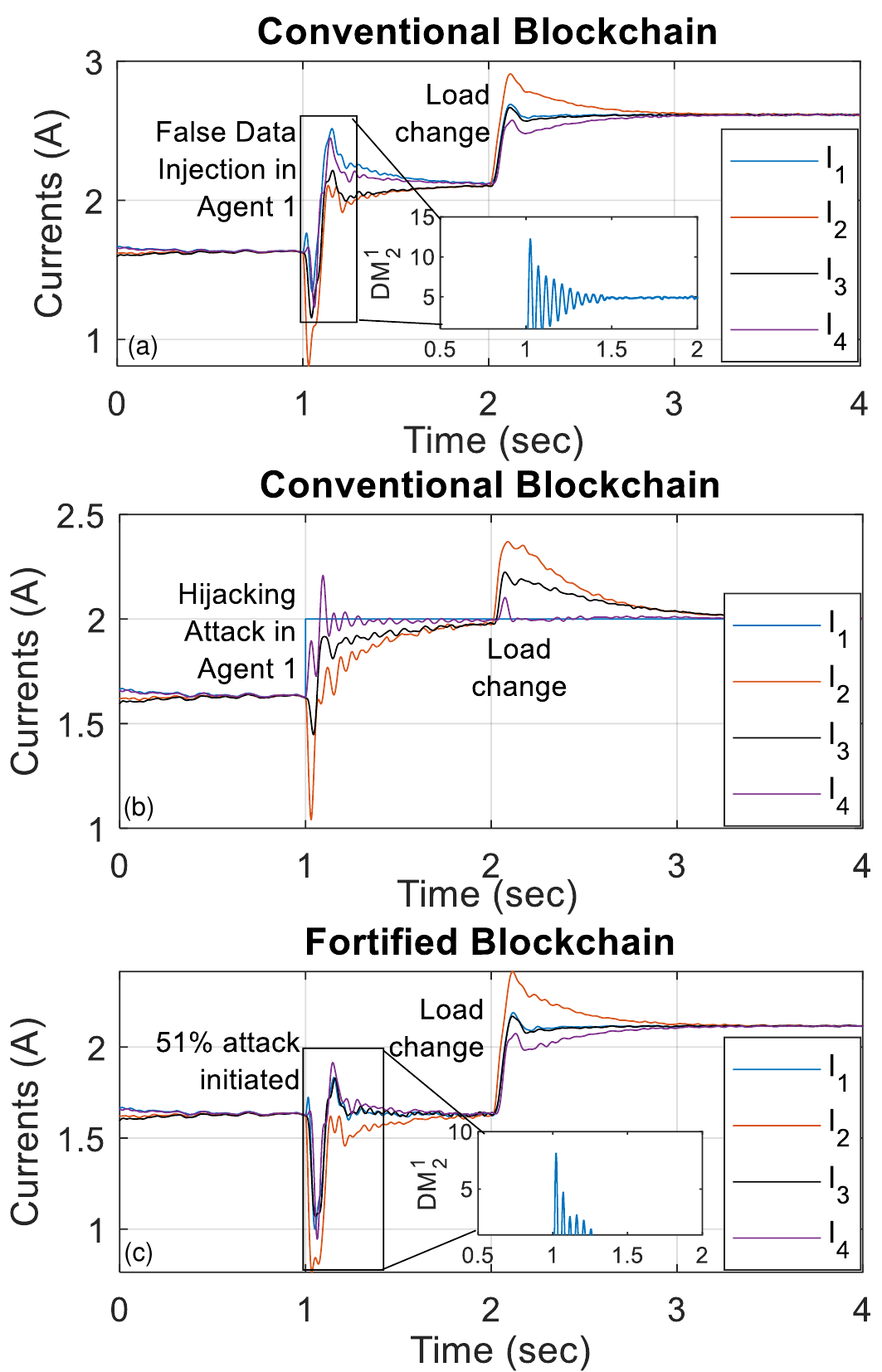}
    \caption{{Performance comparison of the microgrid-Blockchain network with and without additional fortifications as depicted in Table \ref{tab:2}. The conventional Blockchain fails to identify controller-level manipulations and accepts the attacker's manipulations as true data.}}
    \label{fig:51_percent}
\end{figure}

\subsection{\rev{Mitigation of DoS attacks and time delays}}
\begin{figure*}
    \centering
    \includegraphics[width=.7\linewidth]{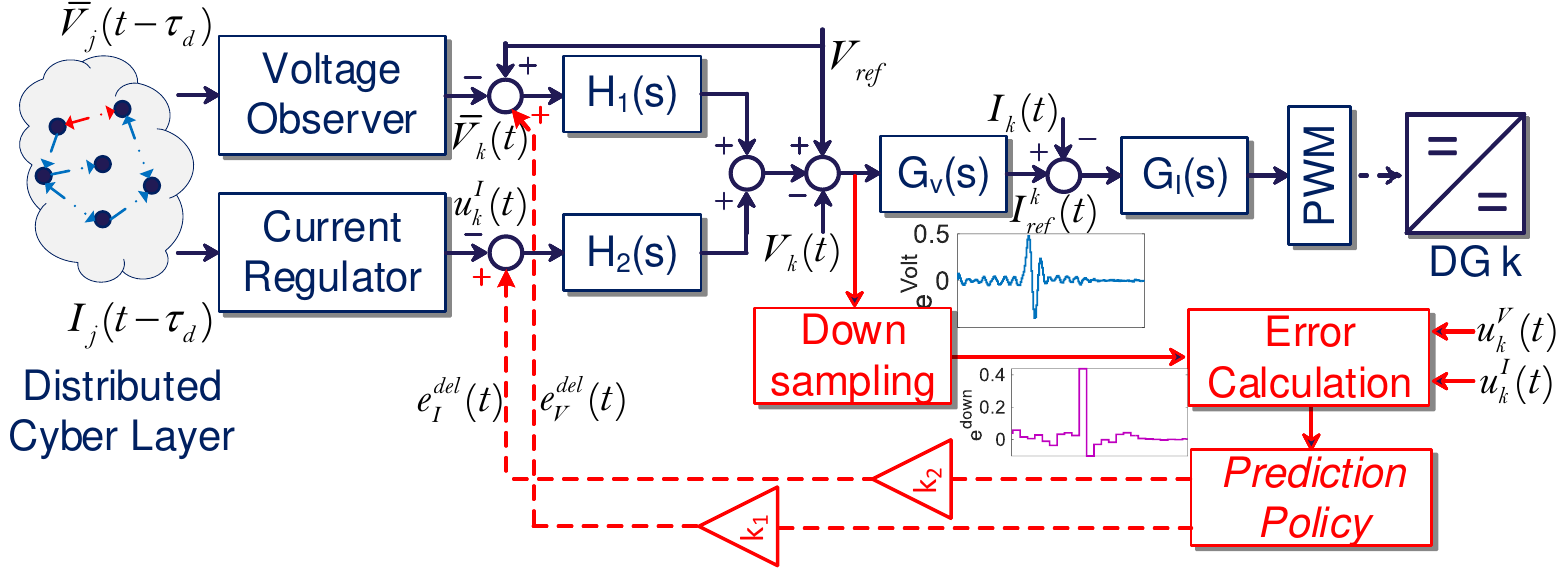}
    \caption{\rev{A physics-based model-free predictive strategy that is augmented into the Blockchain-enabled DC microgrid to mitigate DoS attacks and potential time/computational delays \cite{apec}.}}
    \label{fig:predictive}
\end{figure*}

\rev{A significant advantage of the proposed set of physics-based detection metrics is their ability to also identify denial-of-service (DoS) attacks and network latencies as they would lead to mismatches in measurement values as well.
However, the self-healing strategy as depicted in the preceding subsection may not be able to mitigate such network issues/attacks as the information sharing (via communication links) would be blocked in such cases.
To mitigate these attacks a local compensation framework is established by integrating a prediction policy with the self-healing strategy. This supplementary framework reconstructs an expected version of the attacked signal using a model-free compensatory mechanism as presented in \cite{apec} and depicted in Fig. \ref{fig:predictive}.}

\begin{figure}
    \centering
    \includegraphics[width=\linewidth]{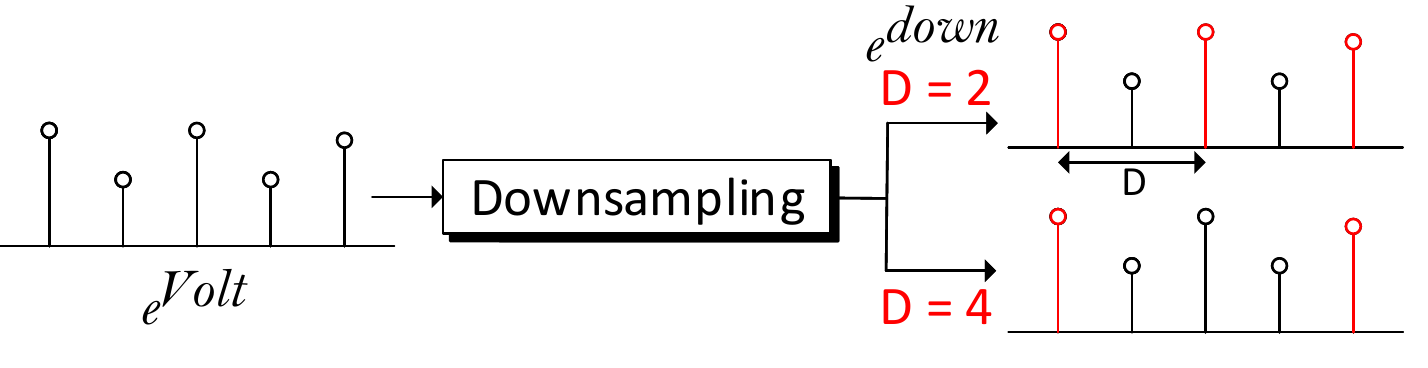}
    \caption{\rev{Downsampling of an error signal into decimated output signals of lower resolution \cite{apec}.}}
    \label{fig:downsampling}
\end{figure}

\rev{In situations where DoS attacks and/or random time delays are encountered, the predictive policy exploits the recipient node's access to the last trustworthy measurement received (just before the occurrence of the \textit{event}). In this case, the measurement not received/delayed (represented as $x(t-d)$) can be approximated using the proposed model-free prediction policy that uses the PI consensuability law \cite{carli2008pi} to generate control signals in the presence of the attack vector. Since the vector seeks to manipulate the flow of measurement and control values in the secondary control layer, the signal representing the error margin $e_k^{Volt}(t)$ for the voltage control loop is downsampled to a reduced value $e_k^{down}(t)$ as per the following formula:
}
\begin{equation}
    e_k^{down} = \sum_{b=0}^{B-1} e_k^{Volt}|nD-b|.h|b|
\end{equation}
\rev{where $h|b|$ represents an impulse response signal whose window length is $B$ downsampling factor is $D$. Downsampling essentially reduces the resolution of the input error signal by decimating it by $D$ samples. This ensures that the dynamic performance of the prior is appropriately matched. A pictorial representation of this process is shown in Fig. \ref{fig:downsampling}. In the figure, the resolution of the error signal $e^{Volt}$ is scaled down by downsampling it into two different signals having resolutions of 2 and 4.}

\rev{For an affirmation of the presence of the attack vector (and/or random time delays), local control, and measurement values are compared with the downsampled signal $e_k^{down}(t)$. Post-comparison, the proposed prediction strategy reconstructs the expected delay compensation signals $e_k(t_k)={e_k^V(t_k), e_k^I(t_k)}$ at the local node side on the basis of the following condition:}
\begin{equation}
    e_k(t_k) = e_k^{down}.[1 1] - {u_k}
\end{equation}
\rev{After reconstruction, the error signal is sent to the prediction policy stage where signal reconstruction is performed again if the attack persists for an increased time period. A similar reconstruction is also performed in the case of large time delays. The overall prediction criterion is provided below.}
\begin{equation}
    ||e_k(t_k)||>\alpha ||exp(-t/T)e_k^{Volt}.[1 1]||
\end{equation}
\rev{where $T(=K_P/K_i)$ represents the time instant associated with the PI control loops, and  represents a tunable parameter. If the condition defined above is satisfied, a trigger is generated that reconstructs $e_k(t_k)$ through a \texttt{Sample} and \texttt{Hold} process with the triggering instant $t_k$. After this, the reconstructed signals (compensators for the expected signal) are provided as inputs to the local secondary (voltage and current) control loops using a set of tunable gains, $k_1$ and $k_2$. The inputs can be defined as follows.}
\begin{equation}
    e_V^{del}(t_k)=k_1e_k(t_k)
\end{equation}
\begin{equation}
    e_I^{del}(t_k)=k_1e_k(t_k)
\end{equation}
\rev{As depicted in Fig. \ref{fig:predictive}, these inputs are fed back into the secondary controller as per the following equations.}
\begin{equation}
    u_k^{Vf}(t)=u_k^V(t)+e_V^{del}(t_k)
\end{equation}
\begin{equation}
    u_k^{If}(t)=u_k^I(t)+e_I^{del}(t_k)
\end{equation}
\rev{where $u_k^{Vf}$ and $u_k^{If}$ represent the final inputs to the secondary control loops as shown in Fig. \ref{fig:predictive}. The proposed strategy effectively handles continued DoS attacks and large-magnitude time delays. Additionally, the error computation framework is used to validate the interruptions in a robust manner.}

\section{{Performance Analysis \& { Simulation} Results}}
{To demonstrate the action of the supplementary detection metric, we validate our results on the considered system in Fig. \ref{fig:sys_model}. \rev{The test system used for strategy validation is developed in the MATLAB R2020a environment.}
\rev{The model (as depicted in Fig. \ref{fig:sys_model}), has four converters (each of 10 kW rating) that are connected to each other through tie-lines $R_i$.  The controller gains are also identical for each converter. The system and control parameters used for the simulation can be found in Table \ref{tab:A1}.}
In the first case study, Fig. \ref{fig:51_percent} presents the magnitude of current signals in {a DC microgrid consisting of $N = 4$ agents (as shown in Fig. \ref{fig:sys_model})}, where a malicious hacker with controller level access executes various types of consensus-disrupting attacks to perform: (i) hijacking, and (ii) false data injection. To perform the hijacking attack, \rev{the attacker activates an attack vector that impairs the iterative update rule leading to an arbitrary behavior. This is simulated through the swapping of a \texttt{True} measured signal with a new (\texttt{False}) constant value, which becomes the reference for other DGs in the system \cite{sahoo2019distributed}. Consequently, all the DGs start operating erroneously resulting in a biased, arbitrary solution. To validate the replaced signal,} the perpetrator interrupts the communication between a claimant (here, an incoming block of data) and the verifier (here, one or more nodes in the Blockchain network) in order to change the authentication decision. Thus, it cons the recipient into believing that a malicious, foreign block is authentic.
\rev{The mathematical model of the hijacking attack can be formulated as
\begin{equation}
    x_j^a(t)=(1-\zeta)x_j(t)+c_j^a
\end{equation}
where, $x_j^a$ represents the final local measurement value from the neighboring nodes, and $c_j^a$ is an attack element. $\zeta$ is a variable representing the presence of the attack vector which only accepts binary values 1 (during the attack), or 0 (otherwise). This leads to a disruption in the behavior of the consensus theory imposing a restriction on $x_j^a(t)$ that can only update during future iterations. This results in the creation of a random steady-state signal value for each individual node that ceases to follow the consensus theory.
}
\rev{To perform} the false data injection attack, \rev{the attacker activates an attack vector that adds an additional (exogenic) signal to the general consensus-based control update during each iteration. This leads to the iterations converging to a feasible but manipulated (\texttt{False}) value, where all the operational states remain confined to their general bounds. To validate the manipulated measurement/control signals,} the attacker internally overwrites any unfavorable decision from the verifier to mask malicious injections into the communication channel.
\rev{The mathematical model of the FDI attack can be described as:
\begin{equation}
    x_j^a(t)=x_j(t)+c_j^a
\end{equation}
The above-mentioned equation allows the transmitted signal to be updated as the attacked version of the signal is still determined based on the time-dependent term $x_j(t)$.
}
Clearly, the non-detection of such consensus-manipulating attacks can have negative repercussions on the system. However, as shown in \rev{Fig.} \ref{fig:51_percent}, a Blockchain network fortified with the proposed set of detection metrics in Table \ref{tab:2} can immediately recognize the presence of {attack elements in the compromised nodes} and hence, effectively negate their actions (through signal reconstruction).}

\begin{table}
	\renewcommand{\arraystretch}{1.3}
	\caption{\rev{System \& Control Parameters of System in Fig. \ref{fig:sys_model}}}
	\label{tab:A1}
	\centering
	\begin{tabular}{c|c|c|c|c|c}
		\hline
		Parameter & Value & Parameter & Value & Parameter & Value \\
		\hline
		\hline
		
		$R_{1}$ & 1.5 $\Omega$ & $R_{2}$ & 1.2 $\Omega$ & $R_{3}$ & 0.8 $\Omega$ \\
		\hline
		
		$R_{4}$ & 0.3 $\Omega$ & $R_{5}$ & 0.5 $\Omega$ & $R_{6}$, $R_{10}$ & 0.6 $\Omega$ \\
		\hline

            $R_{7}$ & 0.45 $\Omega$ & $R_{8}$, $R_{9}$ & 0.4 $\Omega$ & $L_{se_{i}}$ & 3 mH \\
		\hline

            $C_{dc_{i}}$ & 250 $\mu$F & $I_{dc_{min}}$ & 0 A & $I_{dc_{max}}$ & 28 A \\
		\hline

            $V_{dc_{min}}$ & 270 V & $V_{dc_{max}}$ & 360 V & $V_{\rm ref}$ & 315 \\
		\hline

            $I_{\rm ref}$ & 0 & $K^{v}_{P}$ & 5 & $K^{v}_{I}$ & 100 \\
		\hline

            $K^{i}_{P}$ & 2.5 & $K^{i}_{I}$ & 0.05 & $h_i$ & 2.5 \\
		\hline

	\end{tabular}
\end{table}

\begin{table}
	\renewcommand{\arraystretch}{1.3}
	\caption{\rev{System \& Control Parameters of System in Fig. \ref{fig:self-3}}}
	\label{tab:A2}
	\centering
	\begin{tabular}{c|c|c|c|c|c}
		\hline
		Parameter & Value & Parameter & Value & Parameter & Value \\
		\hline
		\hline
		
		$R_{1}$ & 0.2 $\Omega$ & $R_{2}$ & 0.25 $\Omega$ & $R_{3}$ & 0.32 $\Omega$ \\
		\hline
		
		$R_{4}$ & 0.26 $\Omega$ & $R_{5}$ & 0.45 $\Omega$ & $L_{se_{i}}$ & 2 mH \\
		\hline

            $C_{dc_{i}}$ & 120 $\mu$F & $I_{dc_{min}}$ & 0 A & $I_{dc_{max}}$ & 10 A \\
		\hline

            $V_{dc_{min}}$ & 42 V & $V_{dc_{max}}$ & 54 V & $V_{\rm ref}$ & 48 \\
		\hline

            $I_{\rm ref}$ & 0 & $K^{v}_{P}$ & 5 & $K^{v}_{I}$ & 100 \\
		\hline

            $K^{i}_{P}$ & 2.5 & $K^{i}_{I}$ & 0.05 & $h_i$ & 2.5 \\
		\hline

	\end{tabular}
\end{table}

\begin{figure}
    \centering
    \includegraphics[width=.7\linewidth]{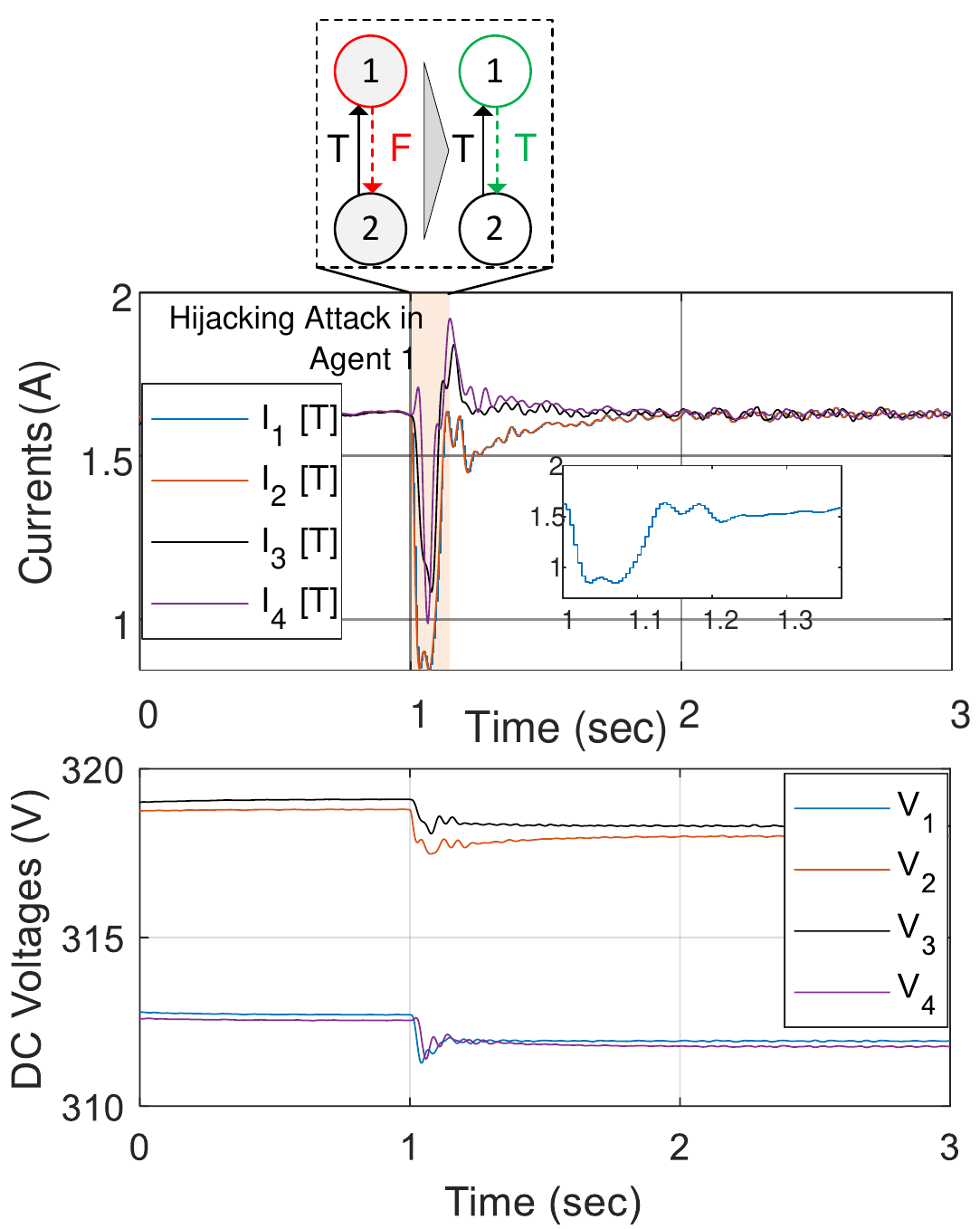}
    \caption{Performance of the self-healing recovery scheme in DLT-based microgrids when the current of agent I is compromised -- in 0.15 sec, the reconstructed signal from agent I is re-transmitted back to its neighbors by retaining the system stability.}
    \label{fig:self-1}
\end{figure}
\begin{figure}
    \centering
    \includegraphics[width=.7\linewidth]{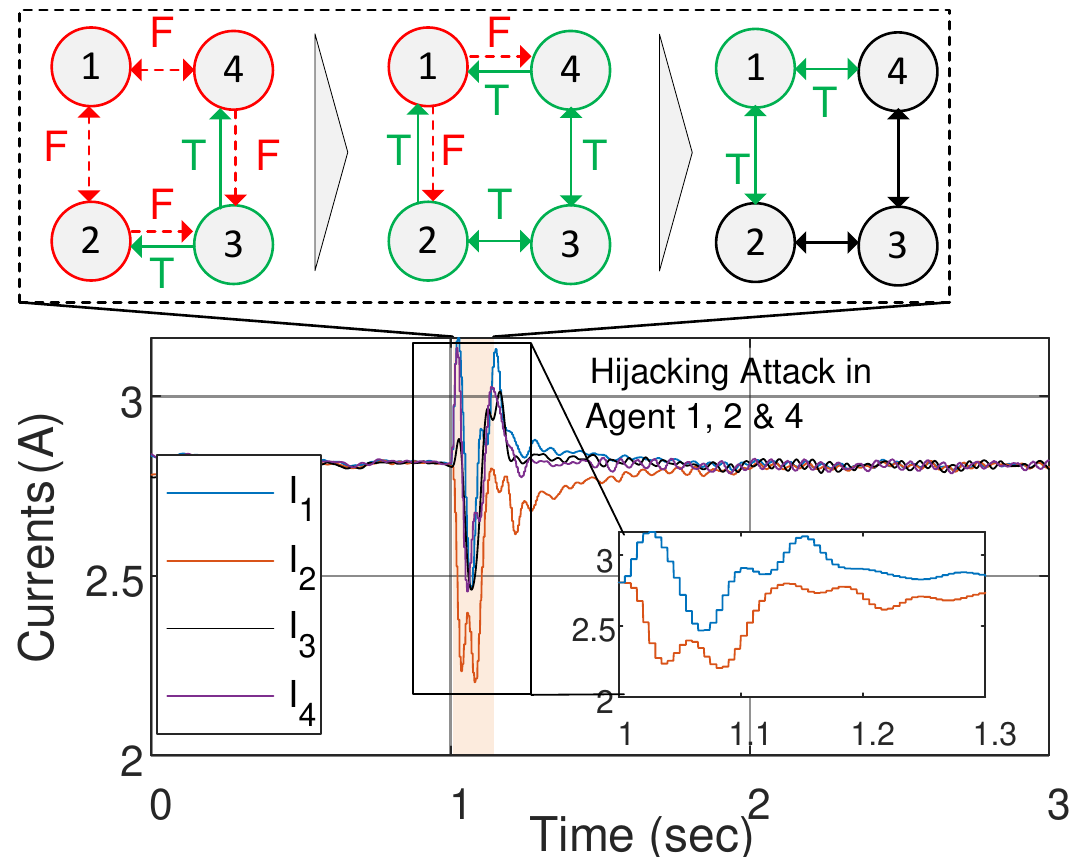}
    \caption{Performance of the self-healing recovery scheme under cyber attacks on $N-1$ converters -- the system is able to heal only from a single trustworthy agent 3.}
    \label{fig:self-2}
\end{figure}
\begin{figure}
    \centering
    \includegraphics[width=.7\linewidth]{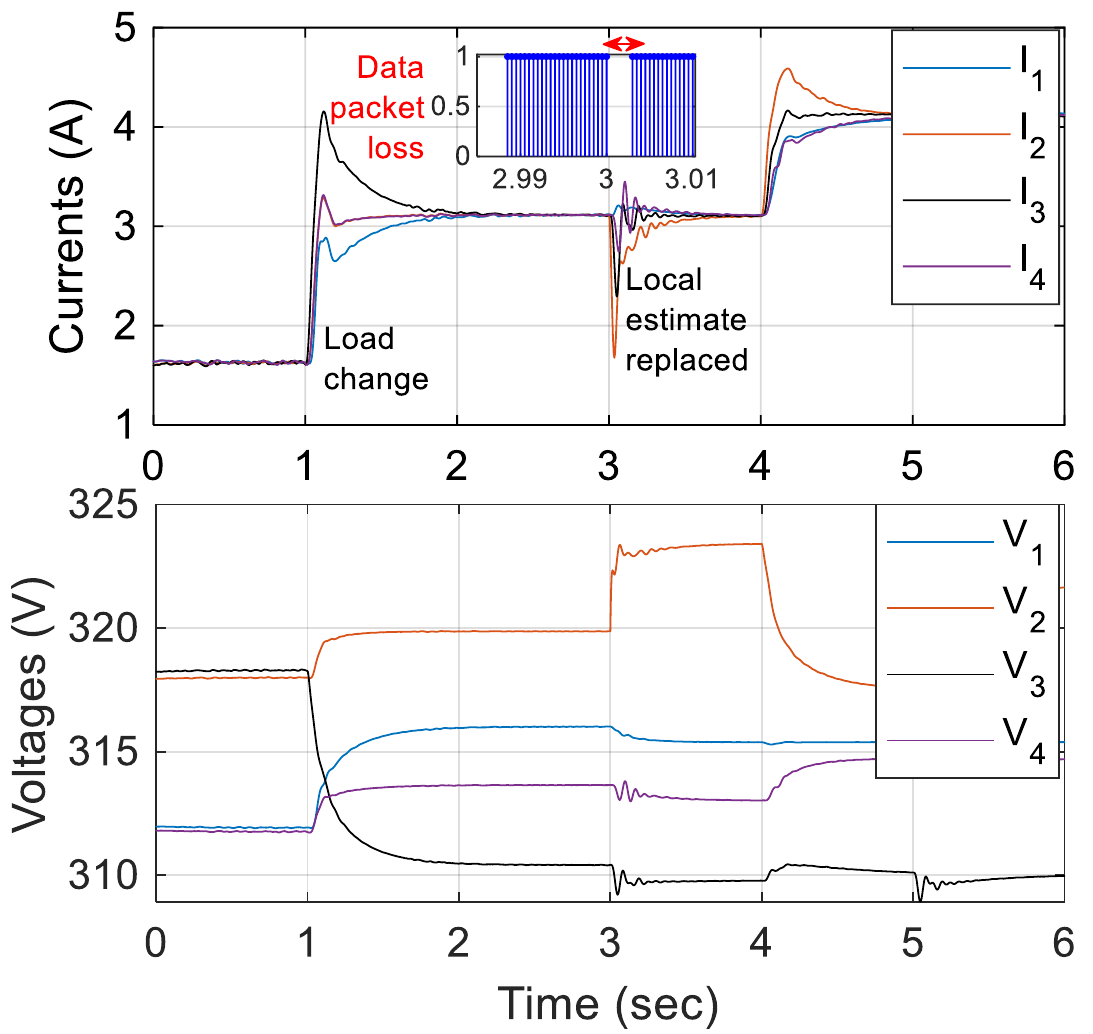}
    \caption{ Performance of the self-healing mechanism during DoS attack when the data packets are lost due to injection of random noise at t = 3 sec -- the local reconstructed estimate \cite{apec} ensures steady-state convergence.}
    \label{fig:dos}
\end{figure}

\begin{figure}
    \centering
    \includegraphics[width=.7\linewidth]{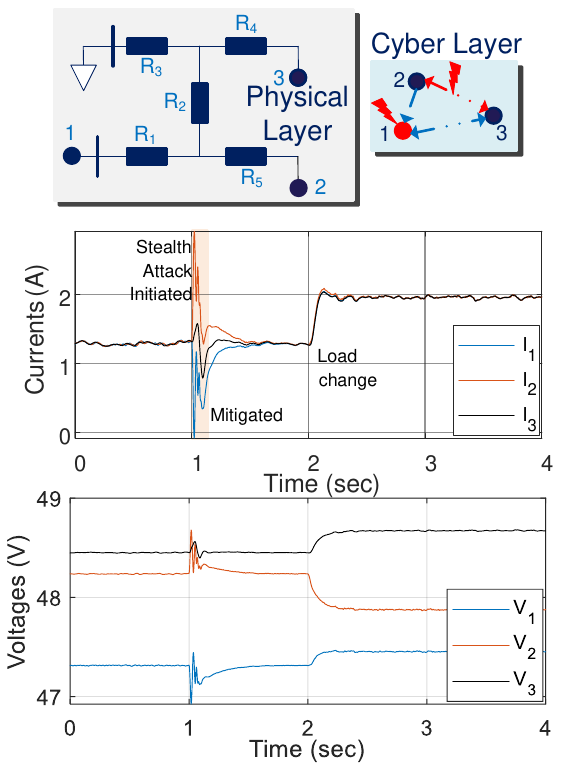}
    \caption{Performance of the self-healing mechanism being independent of the physical parameters of microgrid, where the line resistances and number of converters are different from Fig. 2 -- the system resiliency is still intact under cyber attacks.}
    \label{fig:self-3}
\end{figure}

{To explain the working mechanism of the proposed self-healing strategy in Fig. \ref{fig:Self-healing_1}, we have considered an adversarial situation in the same system, where the attacker has access to the current of agent 1. Fig. \ref{fig:self-1} shows the proposed self-healing Blockchain-based attack detection and mitigation procedure under this scenario. Once $DM_2^1$ detects the presence of attack elements in an agent I, the mitigation mechanism is immediately triggered, which traces a trustworthy current signal from agent 2, as shown in Fig. \ref{fig:self-1}. After acquiring the trustworthy current signal from agent 2, the signal reconstruction is performed to replace the attacked $I_1 (t)$ with a trustworthy event-driven signal $I_1 (t_k)$. As shown in Fig. \ref{fig:self-1}, after the highlighted event-driven signal is replaced, its communication to neighbors is resumed after the validation process from DLT in 0.15 sec. In this way, the considered system operates normally even under a hijacking attack on one of the agents without causing any power interruption. 

To evaluate the $N-1$ resiliency scale of the proposed self-healing strategy, we have also considered the additional adversarial situations in Fig. \ref{fig:self-2}, where the attacker has access to 3 nodes out of 4 agents. In this scenario, we anticipate that the only trustworthy agent III will self-heal the system by broadcasting its trustworthy information in a step-wise manner. When currents of agents 1, 2, and 4 are simultaneously attacked at t = 1 sec in Fig. \ref{fig:self-2}, it can be clearly seen that the system operates normally despite heterogeneous transients from each converter during attacks. The signal authentication is done in a stepwise manner, which has been highlighted in Fig. \ref{fig:self-2}, where the event-driven mitigation is firstly carried out in agents 2 \& 4 and then later followed by agent 1 towards the end. Hence, the resiliency scale of $N-1$ for the proposed strategy is clearly established in a system of $N$ converters.

{ As outlined in Section IV(D), the self-healing mechanism can also be extended to DoS attacks, wherein a local reconstructed estimate designed using physics-governed equations can be replaced by the missing signal for a DoS attack. This can be seen in Fig. \ref{fig:dos}, where a signal interruption can be seen across the link between agent III and IV in the form of 15\% data packet loss. However, the downsampled estimate \cite{apec} obtained from the local error dynamics of the primary controllers of converter III and IV is used to compensate for the delay and achieve steady-state convergence. As a result, it is not dependent on the blockchain-enabled information transmission, since the divergent secondary controller error is substituted locally. It is worth notifying that the self-healing mechanism allows operation under dynamic disturbances, such as load changes (as shown in Fig. \ref{fig:dos}), and line outages as well, which signifies the robustness of this approach.}

{In the final scenario, we consider the performance of the proposed self-healing strategy in another network (shown in Fig. \ref{fig:self-3}) with different system parameters to establish that the strategy is easily scalable to any physical network topology with heterogeneous dynamics.
\rev{This network topology is developed in the MATLAB R2020a environment and has only three converters (each rated 5 kW). Moreover, it functions at different operation conditions as compared to the system utilized for the preceding case study. The system and control parameters for this system can be found in Table \ref{tab:A2}.}

From Fig. \ref{fig:self-3}, it can be clearly established that the performance of the proposed self-healing strategy remains unaltered with respect to any physical network topology or systems exhibiting different dynamics. This can be attributed to the robust physics-informed detection structure in Table \ref{tab:2}, that are independent of structural and temporal dependencies. Based on that, it can be seen that the proposed detection and mitigation strategy augmented into DLT provides resiliency immediately against stealth attacks injected into agent 2 at t = 1 sec. Hence, the security boundaries of DLT is significantly enhanced due to the augmentation of the physics-informed detection metrics and self-healing mitigation strategy for networked microgrids.}

\section{ Experimental Results}}

\begin{figure}
    \centering
    \includegraphics[width=.85\linewidth]{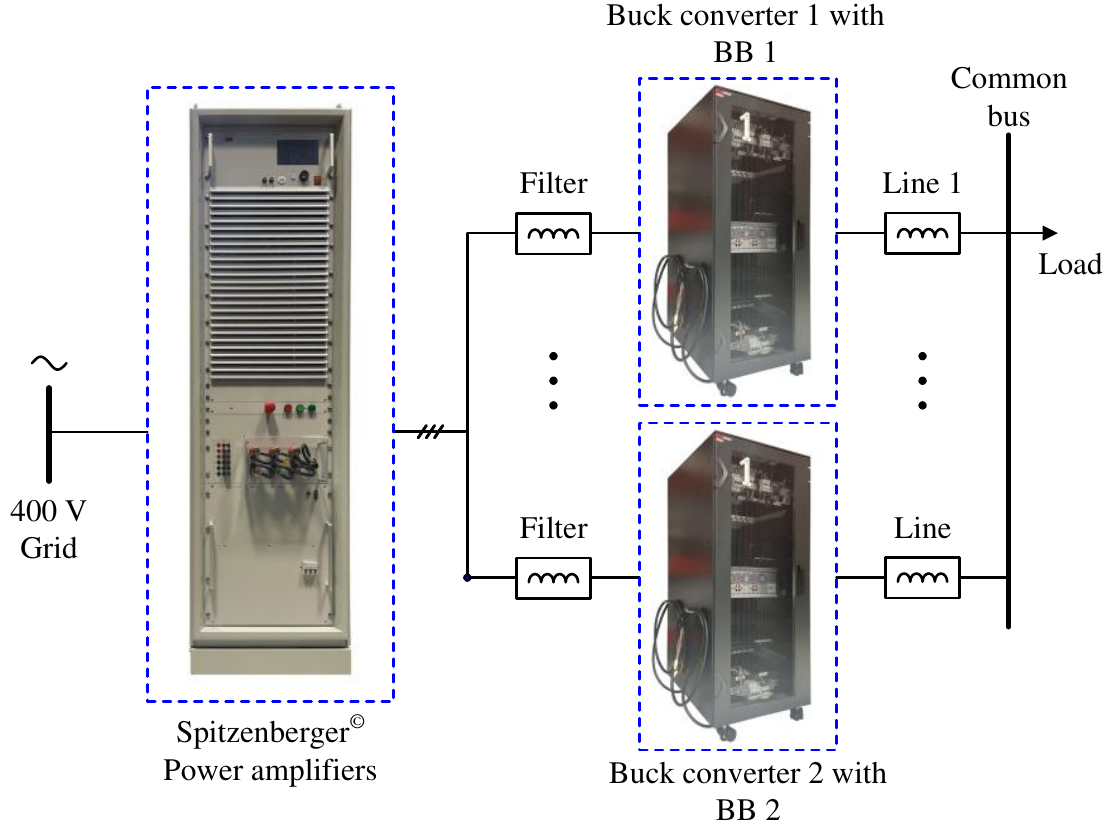}
    \caption{ Experimental setup of a DC microgrid with 2 buck converters controlled by Imperix BoomBox supplying power to the load at common bus. A delay has been programmed before the communication channel to emulate the variable computational time by the blockchain segment for each agent.}
    \label{fig:schematic}
\end{figure}

\begin{figure}
    \centering
    \includegraphics[width=.9\linewidth]{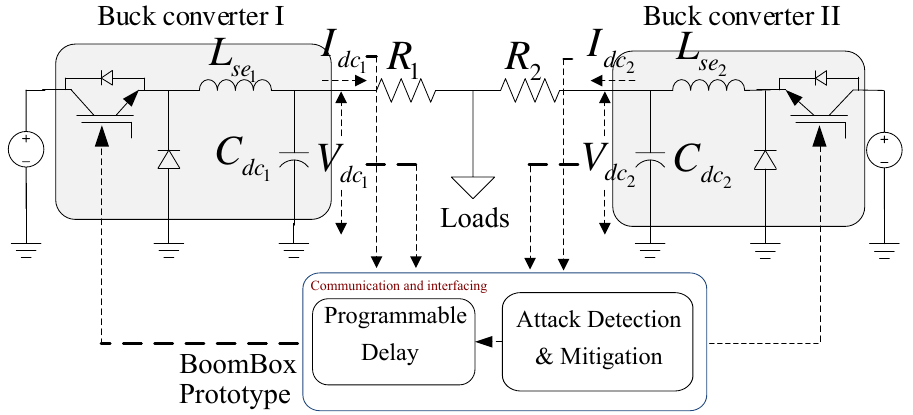}
    \caption{ Single-line diagram of the experimental setup shown in Fig. \ref{fig:schematic}.}
    \label{fig:sld}
\end{figure}

{ The proposed self-healing strategy has been validated in an experimental prototype of DC microgrid shown in Fig. \ref{fig:schematic} operating at a global voltage reference of 48 V with $N$ = 2 DC/DC buck converters. It should be noted that the Spitzenberger power amplifier tied to the Imperix 2 level AC/DC converters have been used as the DC source for both buck converters, equally rated around 7.5 kW. A single-line diagram of the experimental setup can be seen in Fig. \ref{fig:sld}. To emulate the computational \& communication delay equipped with blockchain, a programmable delay blockset available in the BoomBox (BB) is appended before the communication channel. Variable delays and signal interruptions were programmed to emulate real-time cyber-physical operation and robustness of the proposed self-healing mechanism. The system and control parameters can be found in Appendix.}

{ In Fig. \ref{fig:exp}, the performance of the proposed self-healing mechanism is tested under various conditions. In Fig. \ref{fig:exp}(a), a FDIA attack of magnitude 5 A on $I_2$ is carried out. However, due to the proposed self-healing scheme, the reconstructed signal obtained using $I_1$ is immediately substituted to ensure resilience against the cyber attack. This scenario also validates the $N-1$ resiliency scale attribute of the proposed scheme, which has also been simulated in the case study in Fig. \ref{fig:self-1}. Furthermore in Fig. \ref{fig:exp}(b) and (c), the impact of DoS attacks and communication \& computational delay is studied. It can be seen in Fig. \ref{fig:exp}(a) that despite the signal interruption, the current sharing and voltage regulation errors are well-regulated despite the dynamic transients due to the initial mismatch in the reconstruction process. Similarly in Fig. \ref{fig:exp}(c) for a maximum communication delay of 425 ms, the microgrid was initially unstable since the cyber network can only guarantee stability upto a maximum delay of 345 ms. However as soon as the locally downsampled estimate is substituted, the error is regulated under steady-state and dynamics instances. Hence, the performance across the experimental case studies affirm the durability of the proposed self-healing approach under different operating conditions in microgrids.}

\begin{figure}
    \centering
    \includegraphics[width=.95\linewidth]{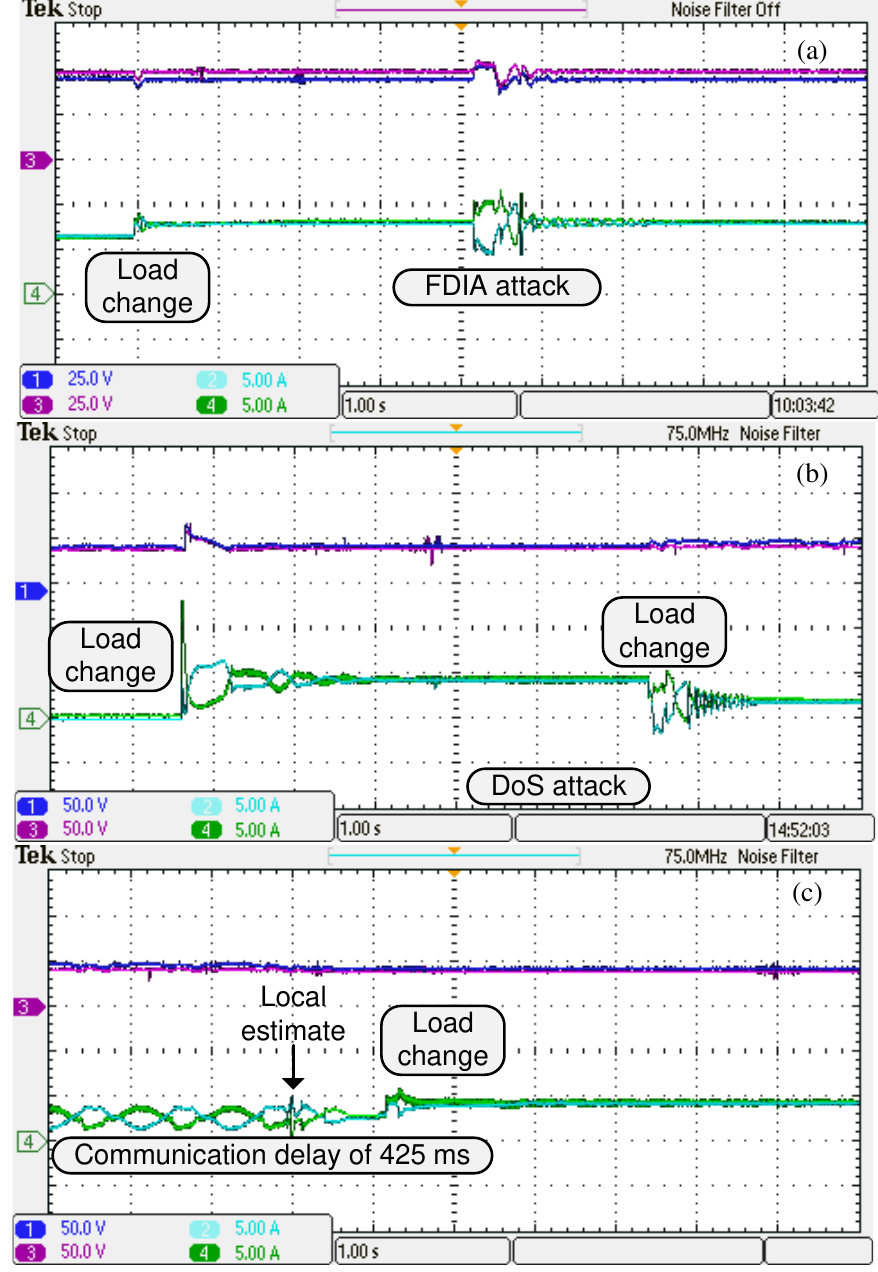}
    \caption{ Performance of the self-healing recovery scheme in DLT-based microgrids under: (a) FDIA attacks, (b) DoS attack, (c) communication delay of 425 ms, which is compensated locally using the downsampled estimate \cite{apec}.}
    \label{fig:exp}
\end{figure}
\section{Conclusion}
{This paper presents a strategy to fortify DC microgrids against potential cyber-attacks through a Blockchain-based approach. The proposed strategy uses a consensus-based verification technique to determine the authenticity of blocks where the majority of the nodes must agree upon the validity of their hash indices for them to be accepted by the recipient node. However, this verification technique may be vulnerable to hijacking attacks, wherein an attacker uses its system-level access to mine malicious, new blocks and fool the recipients into believing their authenticity. Moreover, this access can also be used to fake the consent of the nodes for rejecting authentic blocks or accepting fake blocks. 

To remove this limitation, this paper combines the general features of the Blockchain with unique physics-informed attack detection metrics to detect these attacks and trigger mitigation in their presence. This detection strategy is in the form of a series of rules (threshold criteria) which, when violated, lead to the alteration of a detection metric, signifying attack detection. Detection of any false block in the network leads to its immediate rejection by the recipient. It triggers the activation of a self-healing strategy, where the lost signal/data-point is reconstructed by obtaining previous values of the same signal and current values of neighboring measurements (i.e., sensor inputs from other nodes) from the ledger. This leads to the preservation of system stability even under adverse scenarios, provided that at least one node in the network is trustworthy)}.
\rev{In addition to this, we also provide a strategy for model-free control that is integrated with the proposed self-healing strategy for defense against random time delays and potential DoS attacks. This ensures that any computational delay created due to the Blockchain-based setup is also mitigated in a robust manner.}
{Clearly, a major limitation of the mitigation method proposed in this paper would be its inability to reconstruct a trustworthy version of the signal if all nodes are attacked simultaneously \rev{during FDI attacks}. Hence, we plan to extend this aspect as a future scope of work by overcoming this limitation and providing full resiliency.}

\ifCLASSOPTIONcaptionsoff
  \newpage
\fi
\section*{Appendix}
{ The considered experimental prototype in Fig. \ref{fig:schematic} \& \ref{fig:sld} consists of two DC/DC buck converters rated equally for 7.5 kW. It should be noted that the controller gains are consistent for each converter.\\
\textbf{Plant}: $R_1$ = 0.9 $\Omega$, $R_2$ = 1.2 $\Omega$\\
\textbf{Converter}: $L_{se_{i}}$= 3 mH, $C_{dc_{i}}$ = 100 $\mu$F\\
\textbf{Controller}: $V_{\rm ref}$= 48 V, $K_P^{H_1}$ = 1.92, $K_I^{H_1}$ = 15,
$K_P^{H_2}$ = 4.5, $K_I^{H_2}$ = 0.08, h = 1.5, c = 1.4, $\Upsilon_1$ =
0.025, $\Upsilon_2$ = 0.035}
\bibliographystyle{IEEEtran}
\bibliography{ref}

\end{document}